\documentclass[journal=jpclcd,manuscript=letter]{achemso}

\usepackage[version=3]{mhchem} 
\usepackage{hyperref}
\usepackage{textcomp}

\author{C. Zagorec-Marks}
\email{chase.zagorec-marks@colorado.edu}
\affiliation{Department of Physics, University of Colorado, Boulder, CO 80309, USA}
\alsoaffiliation{JILA, National Institute of Standards and Technology and the University of Colorado, Boulder, CO 80309, USA.}
\author{G. S. Kocheril}
\affiliation{Department of Physics, University of Colorado, Boulder, CO 80309}
\alsoaffiliation{JILA, National Institute of Standards and Technology and the University of Colorado, Boulder, CO 80309, USA.}
\author{T. Kieft}
\affiliation{Department of Physics, University of Colorado, Boulder, CO 80309}
\alsoaffiliation{JILA, National Institute of Standards and Technology and the University of Colorado, Boulder, CO 80309, USA.}
\author{O. A. Krohn}
\affiliation{Combustion Research Facility, Sandia National Laboratories, Livermore, CA 94550, USA}
\alsoaffiliation{Department of Physics, University of Colorado, Boulder, CO 80309}
\alsoaffiliation{JILA, National Institute of Standards and Technology and the University of Colorado, Boulder, CO 80309, USA.}
\author{C. Mart\'{i}}
\affiliation{Combustion Research Facility, Sandia National Laboratories, Livermore, CA 94550, USA}
\author{T. P. Softley}
\affiliation{School of Chemistry, University of Birmingham, Edgbaston, B15 2TT, UK}
\author{J. Z\'{a}dor}
\affiliation{Combustion Research Facility, Sandia National Laboratories, Livermore, CA 94550, USA}
\author{H. J. Lewandowski}
\email{lewandoh@colorado.edu}
\affiliation{Department of Physics, University of Colorado, Boulder, CO 80309}
\alsoaffiliation{JILA, National Institute of Standards and Technology and the University of Colorado, Boulder, CO 80309, USA.}

\title[Title]
  {Vibrational Quantum-State-Controlled Reactivity in the \ce{O2+ + C3H4} Reaction}

\abbreviations{IVR,VVET,UHV,REMPI,TOF-MS,PES}

\begin{document}

\begin{tocentry}

\includegraphics[width=5cm,height=5cm]{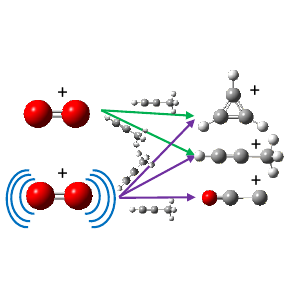}

\end{tocentry}

\begin{abstract}
  Quantum-state-controlled reactivity is a long-standing goal in the field of physical chemistry.
  In this work, we explore the vibrational-state-dependent behavior of the ion-molecule reaction between \ce{O2+} in distinct vibrational states and two isomers of \ce{C3H4}, allene (\ce{H2C3H2}) and propyne (\ce{H3C3H}).
  While most products are formed regardless of the vibrational state of \ce{O2+}, the branching ratios are influenced by vibrational excitation, and a new product, \ce{C2O+}, appears exclusively in the excited-state reactions.
  This selective formation of \ce{C2O+} demonstrates that vibrational excitation can effectively activate a reaction pathway, providing direct evidence of quantum-state control in reactivity.
  These results represent an important step towards the goal of quantum-state-controlled chemistry in molecular systems.
\end{abstract}

Understanding the role of individual quantum states on reactivity is a critical component to developing control of chemical reaction processes, which is an often-stated goal for the physical chemistry community.\cite{Guo2022} 
Original endeavors to achieve this goal investigated the effects of individual quantum states in atom-diatomic reactions and resulted in the development of Polanyi's rules.\cite{Polanyi1972,Polanyi1969,Polanyi1969_2,Pimentel1969}
With these rules, Polanyi suggested that not all forms of energy (vibrational, translational, etc.) are equivalent in determining the progression of a reaction.
Specifically, as observed in these early studies, translational energy facilitated an atom-diatom reaction's progression through an early barrier (i.e., a barrier located along the approach coordinate of reaction); whereas, vibrational energy facilitated progression through a late barrier (i.e., a barrier located along the separation coordinate).
Additionally, it was seen that the energy distribution of the reaction products depended on the position of the barrier, with an early barrier promoting greater partitioning of energy into the products' vibrational modes and a late barrier promoting greater partitioning into the products' translational energy.

Additional experimental support for Polanyi's rules was provided with the advent of lasers. This development renewed interest in state-dependent chemistry, as quantum states of reactants could be directly populated in atom-diatomic reaction systems.\cite{Smith1979}
However, as studies expanded to polyatomic systems, it was found that Polanyi's rules were often not sufficient, particularly for ion-molecule reactions in which submerged barriers are commonly present.\cite{Tanaka1976,Tanaka1979,Anderson1991,Liu2007,Viggiano2001,Manthe2022} 
More recent theoretical efforts have extended Polanyi's rules to be more applicable to polyatomic reactions through the use of quasi-classical trajectory simulations, quantum dynamics calculations, and sudden vector projection models. \cite{Guo2013,czako2021,roncero2013,Zhang2012,Guo2016,Seakins2012,Lendvay2021}

Although there has been a significant amount of theoretical progress in exploring state-dependent reactivity,\cite{Guo2013,czako2021,roncero2013,Zhang2012,Guo2016,Seakins2012,Lendvay2021,Polanyi1972,Polanyi1969,Polanyi1969_2,Pimentel1969,Guo2022} there has been less experimental progress.\cite{Morris1996,Roueff2010,Ng2013,Seakins2012,Schlemmer2020,Crim2008,Anderson2000,Merkt2024,Odom2023,Hudson2017,Hudson2018}
This is due to the experimental difficulty in preparing reactants in a pure quantum state and ensuring that these reactants do not decay to their ground state or, in the case of polyatomic molecules, disperse the excitation energy via intramolecular vibrational redistribution (IVR) prior to the reaction.

Ion–molecule reactions are especially well suited for studies of quantum-state–controlled chemistry, since they are typically exothermic and barrierless. Such reactions proceed rapidly ($k \sim 10^{-9},\space\text{cm}^3$ $\text{s}^{-1}$)\cite{Armentrout2004} and often occur too quickly for significant redistribution of energy among different internal states.
In particular, homonuclear diatomic ions are ideal for studying vibrational-state-dependent reactivity because their lack of an electric dipole moment can result in long-lived, excited vibrational states, and they are immune to IVR by having only one fundamental vibrational mode.\cite{Beckel1979,Dubost1994}
\ce{O2+} is an apt candidate for such studies, as the vibrational lifetime of the $X^{2}\Pi_{g}$ state has been calculated to be as long as $5.7\times10^{5}$ s for the $v=30$ vibrational mode, with even longer lifetimes predicted for lower-lying states, as their decay proceeds solely through quadrupole emission.\cite{Dalgarno1984}

Previously, we investigated reactions of \ce{O2+} in the $X^{2}\Pi_{g}$ vibrationally excited states (\textit{v} = 2,3) with the \ce{C3H4} isomers, allene (\ce{H2C3H2}) and propyne (\ce{H3C3H}), to elucidate the influence of isomeric structure on reactivity. These studies were conducted as a follow-up to our earlier work on the reactions of \ce{C2H2+} with these \ce{C3H4} isomers.\cite{lewandowski2020,lewandowski2021}
While many of the products observed in the original study with \ce{C2H2+} were also detected in the reaction with \ce{O2+} (the primary products \ce{c-C3H3+} and \ce{C3H4+}, and the secondary products \ce{C6H5+} and \ce{C6H7+}), there were also distinct results.\cite{lewandowski2024}
Specifically, it was observed that, contrary to prior studies, both \ce{C3H4} isomers formed a covalently-bound reaction complex with \ce{O2+} and the minor product, which was suggestive of complex formation and had a mass-to-charge ratio (\textit{m/z}) of $40$ \textit{m/z}, was found to be non-reactive to \ce{C3H4}, which was evidence that it may be a product other than \ce{C3H4+}.
\ce{C2O+} was previously proposed as the identity of this non-reactive $40$ \textit{m/z} based on the lack of reactivity and the reaction's atomic composition.
The possible production of \ce{C2O+} in that study, in addition to the knowledge that \ce{O2+} had been formed in a mixture of vibrationally excited states (\textit{v} = 2,3), encouraged our further exploration into how vibrational excitation may influence this reaction.
Here, we explored the effects of vibrational excitation of \ce{O2+} on the ion-molecule reactions, shown below, between \ce{O2+} and the \ce{C3H4} isomers.

\begin{equation}
    \ce{O2+} (v=0) + \ce{C3H4} \xrightarrow{}? 
\end{equation}
\begin{equation}
    \ce{O2+} (v=2,3)+ \ce{C3H4} \xrightarrow{}? 
\end{equation}

We found that the vibrational state of \ce{O2+} governs the formation of the non-reacting $40$ \textit{m/z} product, which is identified as \ce{C2O+}, and this product is produced only upon vibrational excitation of \ce{O2+} despite the presence of a barrierless formation pathway even in the ground state. Thus, we have been able to demonstrate how reactions of \ce{O2+} in different vibrational states can impact the dynamics and ultimately the reaction products. 

The remainder of this letter is organized as follows.
First, a brief description of the experimental apparatus is provided so that the results can be more readily interpreted.
This is followed by a short discussion of previous experimental results for the reactions of \ce{O2+},\cite{lewandowski2024} in the \textit{v} = (2,3) excited vibrational states, with both \ce{C3H4} isomers, allene (\ce{H2C3H2}) and propyne (\ce{H3C3H}).
Next, new data for the reactions of \ce{O2+}, in the ground vibrational state, with both \ce{C3H4} isomers is presented and discussed.
Finally, these results are compared, thereby demonstrating how \ce{O2+}'s vibrational state affects the progression of these reactions.

The experimental apparatus is summarized here (with full details in the Methods section) to provide context for the results. The experiment uses atomic-physics techniques to achieve single-collision conditions and ion translational temperatures below 10 K, where temperature here refers to a spread in energy and is not representative of a true temperature of a Maxwell-Boltzmann distribution.

To achieve these conditions, laser-cooled \ce{Ca+} ions are confined in a linear, quadrupole ion trap within an ultra-high-vacuum chamber, forming a pseudo-crystalline structure known as a Coulomb crystal, which sympathetically cools co-trapped molecular ions through Coulomb interactions. The \ce{O2+} ions are created by two separate (2+1) resonance enhanced multi-photon ionization (REMPI) schemes, which create ions predominantly in the ground state ($\sim91\%$ in ground state) or a mixture of excited vibrational states (\textit{v} = 2 and 3, $\sim68\%$ and $\sim28\%$ of ions, respectively). \cite{Hanneke2025,Kimura1986,Lee1990,Yue2019} Because the ions are separated by $\sim10$ $\mu$m, this sympathetic cooling acts on only translational motion and does not impact vibrational-state populations. Combined with the ultra-high-vacuum environment, which leads to a low rate ($<1$ Hz) of background gas collisions, this ensures that \ce{O2+} remains in its initial vibrational state until a reactive collision occurs. The vacuum conditions further preclude termolecular processes, so internal energy in reaction complexes can dissipate only through bimolecular dissociation or radiative decay.

Reactions are initiated by introducing neutral \ce{C3H4} into the vacuum system at pressures that result in collision rates on the order of $\sim$ 1 Hz. After \ce{C3H4} is admitted for a set amount of time, the trap contents (which include all charged reactants and products) are extracted into a time-of-flight mass spectrometer. Reaction curves are obtained by repeating this procedure with a range of reaction times.

\begin{figure}
        \centering
        \includegraphics[width=1\linewidth]{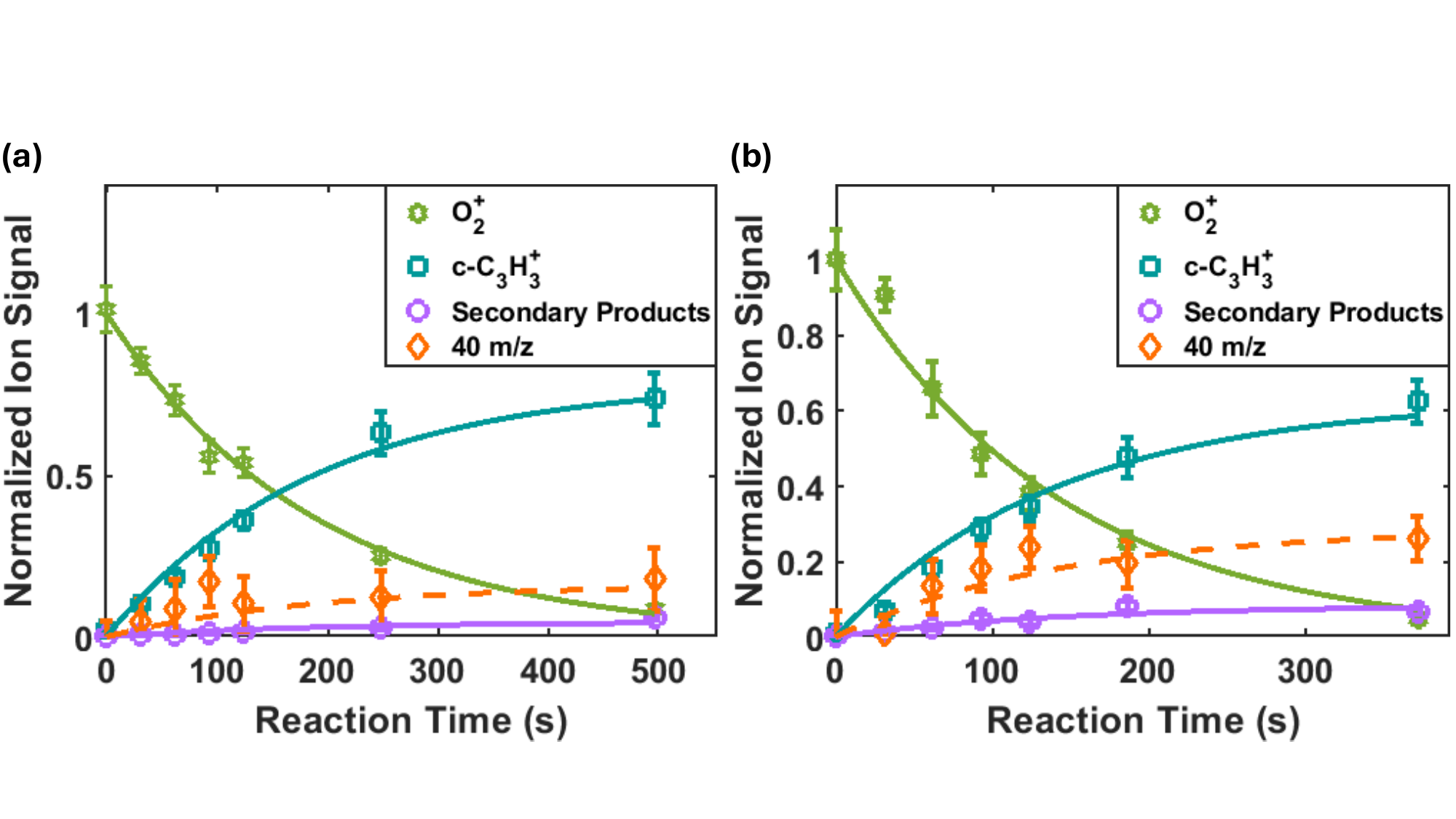}
        \caption{Reaction curves for (a) \ce{O2+}(\textit{v} =2,3) + \ce{H2C3H2} (allene) and (b) \ce{H3C3H} (propyne) reaction series. As the \ce{O2+} signal decreases (green), immediate growth is observed in the $39$ \textit{m/z} (teal) and $40$ \textit{m/z} (orange) mass channels. The product signal in the $40$ \textit{m/z} channel was determined from conservation of charge because of the mass overlap of this product with the trapped \ce{Ca+}. Delayed growth is observed in the $77$ and $79$ \textit{m/z} secondary product channels (combined into one curve displayed in purple). The appearance times, mass-to-charge ratios, and stability of these channels identifies these products as  \ce{c-C3H3+} ($39$ \textit{m/z}), \ce{C2O+}/\ce{C3H4+} ($40$ \textit{m/z}), \ce{C6H5+} ($77$ \textit{m/z}) and \ce{C6H7+} ($79$ \textit{m/z}). \ce{c-C3H3+}, \ce{C2O+}, and \ce{C3H4+} are primary products and the \ce{C6H_y+} ($y=5,7$) ions are secondary products originating from the reaction: \ce{C3H4+}+\ce{C3H4}. Ion signals have been normalized to the fitted initial number of \ce{O2+} ions. Error bars correspond to 1$\sigma$ standard error. Reproduced from Ref. \cite{lewandowski2024} with permission from the Royal Society of Chemistry.}
        \label{fig:v2AlH}
\end{figure}

We first consider previous results where vibrationally excited \ce{O2+}(\textit{v} = 2,3) reacted with two \ce{C3H4} isomers\cite{lewandowski2024}. The reaction curves from that work are shown in Fig. \ref{fig:v2AlH}.
In these reactions, we observed that the primary product formed was \textit{c}-\ce{C3H3+} (blue points), which accounted for $\sim73\%$ ($\sim130$ ions) of all products in the reaction of \ce{O2+} with allene and $\sim62\%$ ($\sim110$ ions) of all products in the reaction of \ce{O2+} with propyne.

Because \textit{c}-\ce{C3H3+} has been observed to be far less reactive than its linear counterpart with a variety of molecules and it has a \textit{m/z}=39,\cite{lewandowski2020,McEwan1994,Ausloos1981,Schlemmer2025,Zhao2014} we assigned this structure to the observed 39 \textit{m/z} product.
In addition to \textit{c}-\ce{C3H3+}, we indirectly observed another product channel at $40$ \textit{m/z} (orange points), which accounted for $\sim17\%$ ($\sim30$ ions) of all products in the reaction of \ce{O2+} with allene and $\sim30\%$ ($\sim70$ ions) of all products in the reaction of \ce{O2+} with propyne.
Although we would have preferred a direct measurement of this product channel, the presence of \ce{Ca+} ($\sim$700 ions) at the same mass led to a large intrinsic uncertainty in the measurement of this much smaller number of product ions. This uncertainty was mostly due to the natural shot-to-shot variations in the loading of \ce{Ca} ions into the trap ($\pm$50 ions), which is comparable in magnitude to the product signal. To reduce the uncertainty arising from the substantial \ce{Ca+} background, we used an indirect method to determine the product ions at 40 \textit{m/z}.

For this indirect method to be valid, two conditions had to be met. First, the total number of ions in the trap needed to remain constant over the course of the reaction. Second, the number of \ce{O2+} ions loaded into the trap also had to remain consistent, with shot-to-shot variations that were small compared to the average number of \ce{O2+} ions loaded ($\sim180$ ions). Once these conditions were satisfied, we determined the number of product ions contributing to the $40$ \textit{m/z} channel using the following equation:

\begin{equation}
    N_{40}(t) = N_{\ce{O2+}}(t=0) - (N_{\ce{O2+}}(t)+N_{\ce{C3H3+}}(t)+N_{\text{Secondary Products}}(t)),
\end{equation}
\noindent where N$_{40}$(t) is the number of ions with a \textit{m/z} of 40 in the trap at time, $t$, that are not \ce{Ca+}, N$_{\ce{O2+}}$(t) and N$_{\ce{C3H3+}}$(t) are the number of \ce{O2+} and \ce{C3H3+} ions in the trap at time \textit{t}, and $N_{\text{Secondary Products}}$(t) is the number of secondary product ions in the trap at time \textit{t} that were produced from the reactions of \ce{C3H4+} with neutral \ce{C3H4}.

Although the 40 \textit{m/z} product could, in principle, have consisted solely of the charge-transfer product \ce{C3H4+}, previous studies have shown that \ce{C3H4+} reacts with 
neutral \ce{C3H4} to form \ce{C6H5+} and \ce{C6H7+},\cite{lewandowski2020,lewandowski2021} 
and we observed both reactive and non-reactive \textemdash\space within the timescale of our experiments \textemdash\space components in the $40$ \textit{m/z} channel besides \ce{Ca+}.
The secondary products from the \ce{C3H4+}+\ce{C3H4} reaction (77 and 79 \textit{m/z}) accounted for only a small portion of the total product signal ($\sim$3\%). This suggested most of the ions with 40 \textit{m/z} were not \ce{C3H4+}, and that an additional species was responsible for the non-reacting fraction.
Given the atomic composition of the reactants, there were only two possible combinations of atoms that could yield \textit{m/z}=40, \ce{C3H4+} and \ce{C2O+}. Because of this, the most reasonable assignment for the non-reactive \textit{m/z}=40 product was \ce{C2O+}.

In these previously studied reactions, it appeared that the non-reacting $40$ \textit{m/z} product (likely \ce{C2O+}) accounted for the majority of the $40$ \textit{m/z} product channel.
Because these reactions were conducted with \ce{O2+} prepared in a mixture of vibrationally excited states, and it is known that vibrational excitations can influence product branching ratios,\cite{Anderson2005} we revisited these reactions with \ce{O2+} prepared in the ground vibrational state to determine how the results of the reaction might be affected by vibrational excitation. The results of these new studies are presented here.

\begin{figure}
        \centering
        \includegraphics[width=1\linewidth]{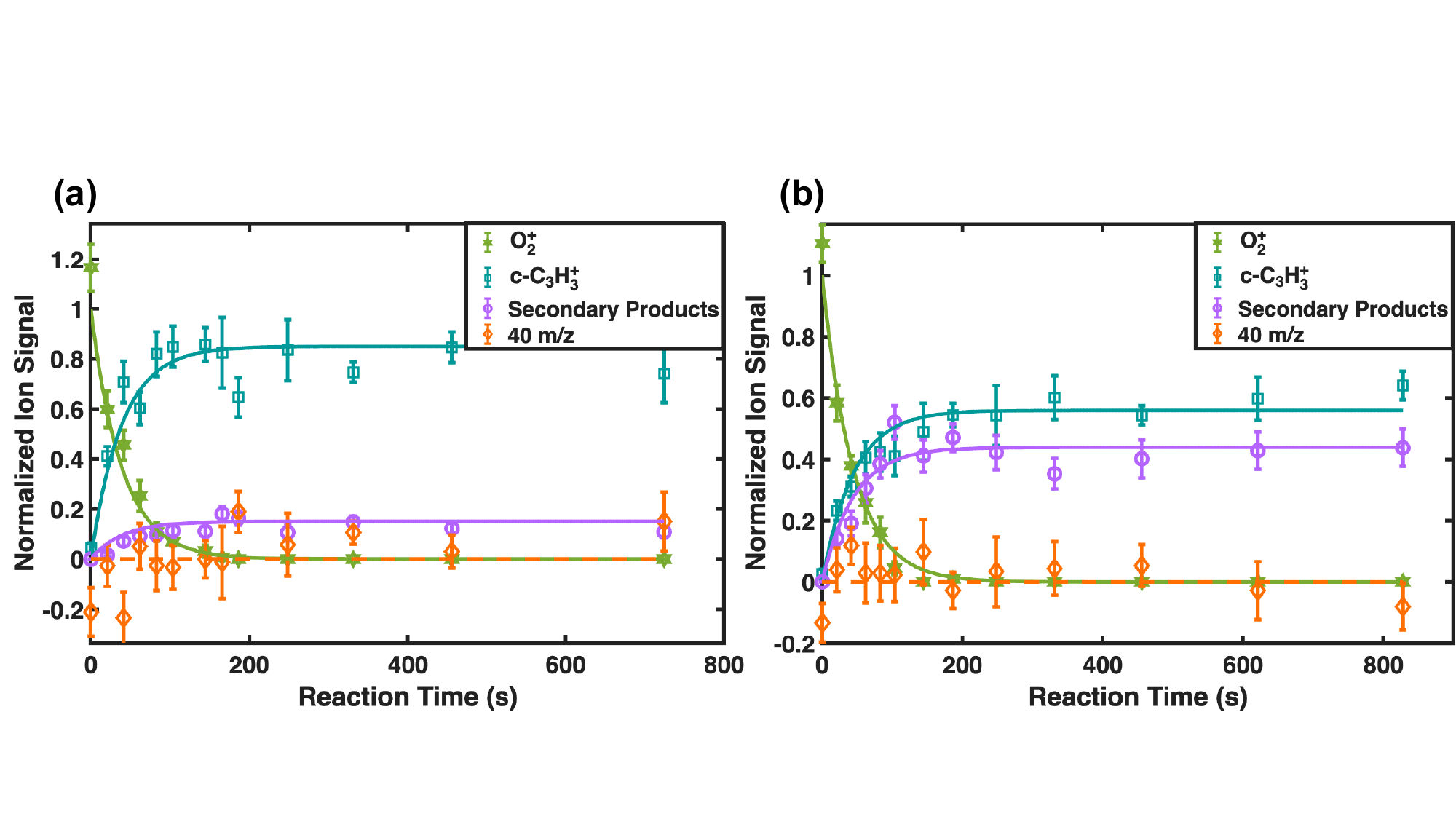}
        \caption{Reaction curves for (a) \ce{O2+}(\textit{v} = 0)  + \ce{H2C3H2} (allene) and (b) \ce{H3C3H} (propyne) reaction series. As the \ce{O2+} signal decreases (green), immediate growth is observed in a non-reacting $39$ \textit{m/z} (teal) channel. Because of the mass overlap of a $40$ \textit{m/z} product with the most abundant isotope of \ce{Ca+}, monitoring of this channel had to be carried out by detection of secondary products produced from the reaction:  \ce{C3H4+ + C3H4} (combined into one curve displayed in purple). A non-reacting product signal in the $40$ \textit{m/z} channel (orange) was estimated from conservation of charge and revealed to be absent in these ground-state reactions. The appearance times, mass-to-charge ratios, and stability of these channels identifies the products formed in the ground-vibrational state \ce{O2+ + C3H4} reactions as  \ce{c-C3H3+} ($39$ \textit{m/z}), \ce{C3H4+},  \ce{C6H5+} ($77$ \textit{m/z}) and \ce{C6H7+} ($79$ \textit{m/z}). \ce{c-C3H3+} and \ce{C3H4+} are the primary products and the \ce{C6H_y+} ($y=5,7$) ions are secondary products. Ion signals have been normalized to the fitted initial number of \ce{O2+} ions. Error bars correspond to 1$\sigma$ standard error.  
} 
        \label{fig:v0}
    \end{figure}

For these studies, we prepared \ce{O2+} ions in the ground vibrational state and reacted them with both \ce{C3H4} isomers. Results from these experiments are shown in Fig. \ref{fig:v0}, where only two primary products were observed.
The majority of product ions from both reactions was the previously observed \textit{c}-\ce{C3H3+} that accounted for $\sim85\%$ ($\sim60$ ions) of all products in the reaction of \ce{O2+} with allene and $\sim56\%$ ($\sim40$ ions) of all products in the reaction of \ce{O2+} with propyne, see Fig. \ref{fig:v0}.
Unlike the reactions with vibrationally excited \ce{O2+}, the only other primary product produced was \ce{C3H4+}, which accounted for $\sim15\%$ ($\sim10$ ions) in the reaction of \ce{O2+} with allene and $\sim44\%$ ($\sim30$ ions) of all products in the reaction of \ce{O2+} with propyne.

\ce{C3H4+} production is determined by measuring the accumulation of products from the \ce{C3H4+} + \ce{C3H4} reaction, which are \ce{C6H5+} and \ce{C6H7+}. These secondary products can be unambiguously detected in our experiment, with no overlap with \ce{Ca+} or any primary reaction products. If we sum the product ions (\textit{c}-\ce{C3H3+}, \ce{C6H5+}, and \ce{C6H7+}) and remaining reactant ions at any point in time along the reaction curve, we see that the sum is equal to the initial number of reactant \ce{O2+} ions, as shown in Supplementary Information Figure 1. This conservation demonstrates that, for ground state \ce{O2+}, no non-reactive $40$ \textit{m/z} products were formed.

This behavior highlights a key distinction between the ground-state and vibrationally excited-state reactions.  
In the vibrationally excited case, a significant fraction of $40$ \textit{m/z} products did not undergo secondary reactions with \ce{C3H4} within the time of the experiment, suggesting that two distinct products may have been present in the $40$ \textit{m/z} channel.
In contrast, in the ground-state case, all $40$ \textit{m/z} products underwent secondary reactions with \ce{C3H4} indicating that this product channel consisted exclusively of a single species, \ce{C3H4+}.
This finding prompted us to confirm the presence of the non-reacting $40$ \textit{m/z} product through a direct measurement.

Although the production of \ce{C2O+} in the vibrationally excited case was previously suggested from charge conservation arguments,\cite{lewandowski2024} we provide additional evidence of the formation of a non-reacting product with $40$ \textit{m/z} here through the use of an isotopically pure \ce{^{44}Ca+} Coulomb crystal and isotopic substitution of the reactants.
To remove the large background signal from \ce{^{40}Ca+} in the mass spectrum, we created an isotopically pure Coulomb crystal of \ce{^{44}Ca+}. This was achieved by first preparing a large Coulomb crystal containing approximately $3000$ \ce{Ca+} ions with the natural isotopic abundance, and then shifting the cooling-laser frequencies to selectively address the \ce{^{44}Ca+} transitions.\cite{Hasegawa2009,Lievens1992} The other \ce{Ca+} isotopes left the trap due to heating, either through deliberate parametric excitation for a particular mass or by lowering the trapping potentials to remove high-energy ions. 
This procedure yielded a small \ce{^{44}Ca+} Coulomb crystal containing approximately 70 ions. While this crystal was sufficiently large to enable product identification, it did not provide enough sympathetic cooling to ensure all product ions remained trapped throughout the full course of a reaction. Consequently, the isotopically purified Coulomb crystal was not used to measure reaction kinetics, but instead served to identify the presence of a non-reacting $40$ \textit{m/z} product.

After preparing the \ce{^{44}Ca+} Coulomb crystal, \ce{O2+} ions were loaded into the trap in a mixture of vibrationally excited states (\textit{v} = 2,3). Reactions were then performed using either deuterated allene (\ce{D2C3D2}, \textit{m/z}=44) or propyne (\ce{D3C3D}, \textit{m/z}=44), which were selected to isolate growth in the $40$ \textit{m/z} channel and thereby support the identification of \ce{C2O+} as a reaction product.

\begin{figure}
        \centering
        \includegraphics[width=1\linewidth]{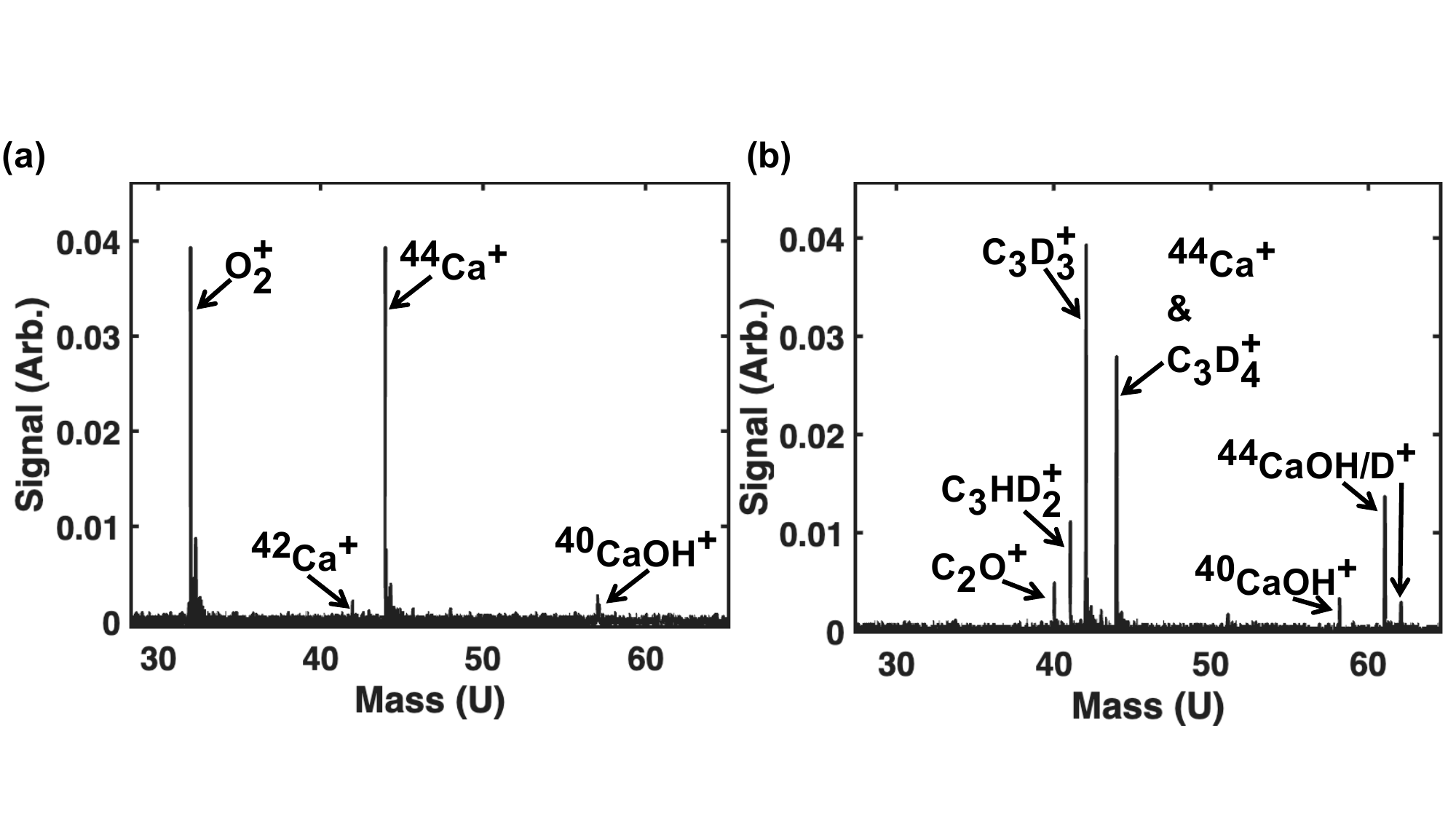}
        \caption{TOF-MS traces depicting the reaction of \ce{O2+}(\textit{v} = 2,3) + \ce{C3D4} within an isotopically pure \ce{^{44}Ca+} Coulomb crystal. These data correspond to a reaction using d4-propyne (\ce{D3C3D}). (a) Example TOF-MS trace at \textit{t} = 0 in which \ce{O2+} ions are observed at $32$ \textit{m/z} and \ce{^{44}Ca+} at $44$ \textit{m/z}. The two smaller peaks present correspond to \ce{^{42}Ca+} and \ce{^{40}CaOH+} ($57$ \textit{m/z}). We observe no ions in the $40$ \textit{m/z} location with single ion sensitivity. (b) Example of a TOF-MS trace at \textit{t} = 300 s in which \ce{O2+} has fully reacted with \ce{C3D4} to form: \ce{C2O+} at $40$ \textit{m/z}, \ce{C3D3+} at $42$ \textit{m/z}, and \ce{C3D4+} at $44$ \textit{m/z}. Additional peaks are also observed at $41$ \textit{m/z} that corresponds to \ce{C3HD2+}, which has formed from H-D swapping between \ce{C3D4+} and ambient water molecules within the chamber, and at $57$ \textit{m/z}, $61$ \textit{m/z} and $62$ \textit{m/z} corresponding to \ce{CaOH+} and \ce{CaOD+} ions, which are formed from background reactions of \ce{Ca+} with ambient water.}
        \label{fig:Ca44}
\end{figure}

Growth was observed in the $40$ \textit{m/z} channel in addition to all other expected products (\ce{c-C3D3+} with \textit{m/z}=42 and \ce{C3D4+} with \textit{m/z}=44) as shown in Fig. \ref{fig:Ca44}. Additional control experiments (e.g., not admitting neutral propyne/allene or not trapping \ce{O2+})  confirmed that the $40$ \textit{m/z} growth originated from reactions between \ce{O2+} and \ce{C3D4} and not via reactions with any contaminants.
Therefore, because growth in the $40$ \textit{m/z} channel was observed only when both \ce{O2+} in vibrationally excited states (\textit{v} = 2,3) and \ce{C3X4} (X = H, D) were present, this signal was assigned to the formation of \ce{C2O+}.

To better understand the origin of \ce{C2O+} production in \textit{only} the vibrationally excited states, a portion of the potential energy surfaces (PESs) for the reactions between ground-state \ce{O2+} and the two \ce{C3H4} isomers were calculated using KinBot to search for barrierless pathways to the observed products \cite{KinBot2020}.

\begin{figure}
        \centering
        \includegraphics[width=1\linewidth]{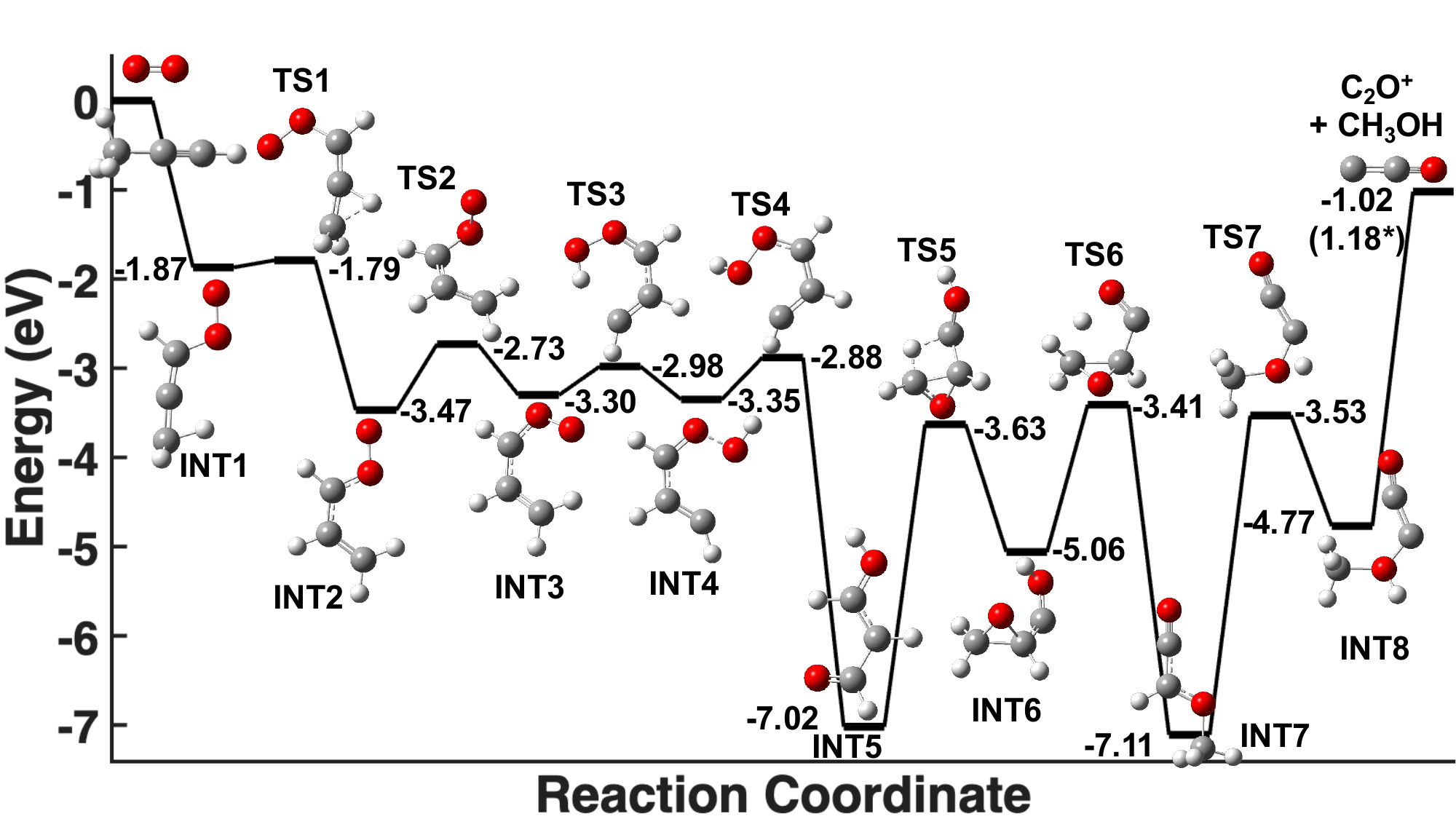}
        \caption{Potential energy surface for the production of \ce{C2O+} in the the reaction between \ce{O2+}(\textit{v} = 0) and propyne (\ce{H3C3H}). All structures were optimized at the MP2/aug-cc-pVTZ level of theory. Single-point energies of these structures were calculated at the CCSD(T)-F12/cc-pVDZ-F12 and have been zero-point energy corrected at the MP2/aug-cc-pVTZ level. All energies are shown relative to reactants energy at infinite separation, with INT$1$ corresponding to the first stationary point following complex formation. Note that two values are reported for the production of \ce{C2O+}, one from CCSD(T)-F12 and the asterisked value computed using values from the ATcT database.\cite{Ruscic2004,Ruscic2005} See Computational Methods for more explanation.}
        \label{fig:C2OPES}
\end{figure}

The portion of the PES that shows a possible pathway from reactants to \ce{C2O+} is shown in Fig. \ref{fig:C2OPES} (propyne) and Supporting Information Figure 2 (allene). The energies of the extrema on the surface were zero-point energy corrected and are displayed relative to the reactants' energy at infinite separation.
It should be noted that, unlike higher-pressure systems where collisional stabilization can remove energy from the reaction complex, the single-collision conditions employed here preclude such energy loss. As a result, the total energy of the complex remains equal to its initial energy, preventing it from becoming trapped in local minima and allowing it to readily traverse any submerged barriers on the potential energy surface.

The propyne surface begins with a moderately exothermic ($1.87$ eV) attachment of \ce{O2+} onto \ce{C3H4} at INT $1$.
After this initial attachment, the complex can traverse the surface sampling various orientations of the initial attachment complex before the \ce{O-O} bond is broken at TS $4$.
The breaking of the \ce{O-O} bond and the progression to INT $5$ is highly exothermic ($\sim4$ eV exothermic relative to TS $4$).
After the \ce{O-O} bond has been severed, the complex can progress through the surface by a series of low-submerged-barrier proton transfers.
These proton transfers result in the complex having rearranged into INT $8$, where the complex dissociates into \ce{C2O+} and the neutral co-product \ce{CH3OH}.

The calculated potential energy surfaces revealed that production of \ce{C2O+} is exothermic and barrierless for the reactions between \ce{O2+}(\textit{v} = 0) and both \ce{C3H4} isomers.
KinBot also discovered over $200$ energetically allowed product combinations (with $22$ unique mass-to-charge ratios), including all of those observed experimentally.
Using just thermodynamic arguments with the results of the KinBot calculations, one would conclude that \ce{C2O+} should be a very minor product as the $127$-th lowest-energy product, and other products that we did not detect experimentally, such as \ce{C2H4+} and \ce{C2H3OH+}, should have been more prevalent.
The observation of \ce{C2O+} and absence of many products predicted to be more abundant demonstrate that quantum dynamical effects must be at play in these reactions.
Such dynamical effects are not uncommon \cite{Ng2014,Lewandowski2025,Wester2017,Hudson2018,Zhang2022}, and have been observed to prevent barrierless reactions from occurring \cite{Ng2014,Lewandowski2025}.

Intuitively, one could argue that production of \ce{C2O+} was facilitated in the excited-state reactions because the internal reaction coordinate at TS $4$ on the \ce{C2O+} formation PES is directed along the \ce{O-O} bond.
However, for this facilitation to occur, vibrational energy had to remain localized in the \ce{O-O} bond long enough for this reaction to take place, i.e., until TS $4$ is reached. 
While vibrational energy can, in principle, be removed from a molecule in a variety of ways, including through collision quenching, radiative decay, or intermolecular energy transfer, several of these are unlikely in this particular system. Collisional quenching of excited vibrational modes is not present because of the ultra-high-vacuum environment of the experiment, which leads to single collision conditions. Radiative decay of vibrational excitation occurs on timescales that are not relevant here; neither \ce{O2+} nor the transient reaction complex has enough time to radiate before the complex dissociates into products. Here, there are two possible modes of energy transfer that could have redistributed the \ce{O2+}'s initial vibrational energy, a vibration-vibration energy transfer (VVET) during the initial collision and intramolecular vibration redistribution (IVR) during complex formation.

VVET is known to be most efficient when the two colliding partners have vibrational energy spacings that are similar.
For example, collisions between vibrationally excited \ce{CO2} and \ce{N2} have been shown to exhibit an order-of-magnitude decrease in energy-transfer efficiency when \ce{^{14}N2} (with an energy mismatch of approximately 18~cm$^{-1}$) is replaced by \ce{^{15}N2} (with an energy mismatch of approximately 97~cm$^{-1}$). \cite{Sharma1969}
For the collisions considered here, the energy of one quanta of the \ce{O-O} stretch in \ce{O2+} ($\tilde\nu$ = 1860~cm$^{-1}$, \textit{v} = 2-1) is substantially mismatched with the nearest energetically accessible vibrational modes of allene (\ce{H2C3H2}) and propyne (\ce{H3C3H}), with energy differences of approximately 433~cm$^{-1}$ and 424~cm$^{-1}$, respectively. \cite{shamir1968,Farooq2014,Shimanouchi1972} Such large energy mismatches make VVET unlikely. Moreover, even if VVET were to occur, the transfer of a single vibrational quantum would not fully quench the excitation of the \ce{O-O} stretch; instead, multiple sequential quanta would need to be transferred to prevent vibrational energy from remaining localized in the bond during the initial collision and complex formation. 

Although multi-quantum VVET has been observed in systems with high vibrational excitation levels ($\textit{v} > 15$),\cite{Wodtke1996} single-quantum transfer processes are nearly an order-of-magnitude more probable for low vibrational excitations ($\textit{v} < 5$).\cite{Guo2020} Consequently, given that the highest excitation involved in the present reactions was \textit{v} = 3, it is reasonable to conclude that, if VVET occurred at all, it was dominated by single-quantum transfer processes. Thus, following the initial collision between \ce{O2+} and \ce{C3H4}, one or two, and potentially even three, vibrational quanta likely remained localized in the \ce{O-O} stretch. A key question then arises: does this vibrational energy remain localized in the \ce{O-O} bond during complex formation, or was it rapidly redistributed throughout the reaction complex via IVR?

An estimate for the efficiency of IVR in this reaction can be made based on estimations of the reaction complex's lifetime and typical times associated with IVR.
The lifetime of the reaction complex can be roughly approximated through the use of statistical models such as Rice-Ramsperger-Kassel-Marcus (RRKM) theory or transition-state theory.\cite{Bohn2014,Svensson1995}
While such calculations assume a thermal distribution of states over the course of the reaction as a result of rapid IVR rates, they can be useful for providing an order-of-magnitude estimate for the complex's lifetime for systems not in thermal equilibrium, as was the case here.

To estimate this lifetime, we performed a simple unimolecular decomposition transition-state theory calculation.
In this calculation, the highest energy transition-state structure (TS$5$, Supporting Information Figure 3) was chosen so as to maximize the complex's estimated lifetime, and the reaction complex's initial structure (INT$1$, Supporting Information Figure 3) was chosen as the pre-decomposition molecule.
These two structures were the only structures used in this calculation.
Using these conditions, the calculated lifetime was estimated to be $\sim1$ ps.
This estimated lifetime is comparable to instances of fast IVR that have been measured to occur over $\sim8$ ps;\cite{Abel2001,Crim2004} however, it has been found that IVR times typically lie in the $20-90$ ps range.\cite{Ebata2005,Zhang2024,Pate2004}
The difference in these time ranges suggests that IVR could not occur fully before the complex dissociated into the final products.
This implies that even in the unlikely case that \ce{O2+} transferred $1$ quanta of vibrational energy during the initial collision with \ce{C3H4}, the leftover vibrational energy remained localized in the \ce{O-O} stretch, as it would be unable to couple into the complex's vibrational modes before dissociation occurred.
It is reasonable to conclude that this localization facilitated the breaking of the \ce{O-O} bond at TS $4$ (Fig \ref{fig:C2OPES}) of the calculated potential energy surface and ultimately led to the formation of \ce{C2O+}.
However, this qualitative explanation only rationalizes the observation of \ce{C2O+} production. This explanation does not clarify why other products that require breaking the O-O bond, such as \ce{C2H3OH+}, are not produced. To understand the full extent of how dynamics influences this particular reaction, one would need to perform quantum dynamics simulations on the complete PES. Such an endeavor is beyond the scope of this work, but we hope our results spur others to explore the dynamics of this complex system.

The results presented here confirm that \ce{C2O+} is a product in the reactions between \ce{O2+}(\textit{v} = 2,3) and \ce{C3H4}, and demonstrate that the production of \ce{C2O+} is governed by dynamical effects rather than just by energetics.
While we presented qualitative arguments for how these dynamical effects may appear, a more quantitative argument based on quasi-classical trajectories, or other methods, is necessary but would be quite challenging to compute based on the number of atoms involved.

Regardless of the complete details of how dynamics direct the reaction complex to one product over another, it is clear from the experimental data that the pathway to \ce{C2O+} is effectively activated by vibrational excitation of \ce{O2+}.
The result that preparation of \ce{O2+} in a vibrationally excited state enables formation of an otherwise unobserved product demonstrates the potential for controlling \ce{C2O+} production in this reaction. In principle, the branching ratio for \ce{C2O+} might be further enhanced, possibly to the point of exclusive formation, by preparing \ce{O2+} in higher vibrational states. 
Exploration of this possibility, as well as the broader role of \ce{O2+} vibrational excitation in other reactions, would be well suited to future studies, particularly given the long lifetimes of the excited vibrational states of \ce{O2+}.
Similar explorations could be made with other homonuclear diatomics and may reveal interesting chemical behaviors and control.

Although full quantum-state control of the \ce{O2+} + \ce{C3H4} reaction has not yet been realized, this work represents a meaningful advance toward that goal. The results highlight substantial opportunities for probing quantum-state–dependent reactivity and point toward a future in which chemical reaction outcomes can be fully controlled at the quantum level.

\section{Methods}
\subsection{Experimental}
Reactions were conducted using our custom-built ion-trapping apparatus, which has been described previously in-depth.\cite{lewandowski2017}
In short, this apparatus consists of a linear, quadrupole ion trap held within an ultra-high-vacuum environment ($\sim10^{-10}$ Torr).
\ce{Ca+} ions were loaded into the trap by non-resonantly photoionizing ($\sim7$ mJ/pulse, 355 nm, 10 Hz) an effusive source of \ce{Ca} in the center of the ion trap.
These trapped \ce{Ca+} ions were then laser cooled  to sub-Kelvin temperatures using the doubled output of a Ti-Sapph laser ($\sim397$ nm, $2$ mW) and a diode laser ($\sim866$ nm, $2$ mW).
\ce{O2+} ions were then co-trapped with the \ce{Ca+} ions by resonantly photoionizing a supersonic molecular beam of \ce{O2} in He ($2\%$ mixture) in the center of the trap.
Two ($2+1$) REMPI schemes at $\sim301$ nm and $\sim288$ nm were used to form \ce{O2+} ions predominantly in the ground or vibrationally excited states (\textit{v} = 2,3), respectively.\cite{Hanneke2025,Kimura1986,Lee1990,Yue2019}

The \ce{O2+} ions were then sympathetically cooled by the laser-cooled \ce{Ca+} ions to secular temperatures $\sim1$ K. As the ions cooled, they self-arranged into a pseudo-crystalline structure known as a Coulomb crystal.
Within the Coulomb crystal, the spacing between ions is around $\sim10$ $\mu$m due to the Coulomb repulsion. Consequently, only the translational motion is sympathetically cooled, while the vibrational and rotational state populations remain unperturbed by the co-trapped ions.

After the \ce{O2+} ions had been trapped and cooled, reactions were initiated by admitting either neutral allene (\ce{H2C3H2}) or propyne (\ce{H3C3H}) into the chamber through a pulsed leak valve at a constant pressure ($\sim1.2\times10^{-9}$ Torr for the ground-state reactions and $\sim 0.3\times10^{-9}$ Torr for the excited-state reactions) for variable periods of time ($0-350$ seconds).
At various reaction times, the contents of the trap were ejected into a time-of-flight mass spectrometer (TOF-MS) to measure the number of ions at each \textit{m/z}.  A new Coulomb crystal was loaded for each measurement. 
Measurements at each time point were repeated $>7$ times, and the mean and standard error of the mean were plotted and globally fit, with the following chemical equations and pseudo-first order rate equations (where allene and propyne are treated in excess), as a function of reaction time to produce the observed reaction curves.

\begin{equation}
    \ce{O2+ + C3H4 ->[\text{k$_{1}$}] C3H3+ + H + O2}
\end{equation}
\begin{equation}
    \ce{O2+ + C3H4 ->[\text{k$_{2}$}] C3H4+ + O2 ->[\ce{C3H4}] \text{Secondary Products}}
\end{equation}

\begin{equation}
    \frac{d(\ce{O2+})}{dt} = -(k_{1}+k_{2})\times\ce{O2+}(t)
\end{equation}
\begin{equation}
    \frac{d(\ce{C3H3+})}{dt} = k_{1}\times\ce{O2+}(t)
\end{equation}
\begin{equation}
    \frac{d(\ce{C3H4+}+\space \text{Secondary Products})}{dt} = k_{2}\times\ce{O2+}(t)
\end{equation}
Where $k_{1}$ and $k_{2}$ correspond to fitted pseudo-first order rate constants for production of \ce{C3H3+} and \ce{C3H4+}, respectively, and are reported in Supplementary Information Table 1.
As \ce{C3H4+} shares a mass channel with the most abundant isotope of \ce{Ca+}, direct observation of \ce{C3H4+} resulted in large uncertainties from the fitting routine originating from fluctuations in \ce{Ca+} loading.
These fluctuations from \ce{Ca+} ($\pm50$ ions) were avoided by indirectly monitoring the \ce{C3H4+} product channel by instead fitting to the directly observed growth of secondary products from the reactions of \ce{C3H4+} with additional neutral \ce{C3H4} molecules (with total fluctuations of $\sim5$ ions).
While this approximation does not capture the short reaction time behavior of \ce{C3H4+}, it captured the \ce{C3H4+} branching in the long reaction time limit.

\subsection{Computational Methods}
Our goal in this work was to identify feasible pathways to the observed products and not to provide a complete and accurate kinetic characterization of the processes at play, especially given the experimental results that suggest dynamics dominate the branching between the open channels. 

To identify feasible pathways, potential energy surfaces (PES) for the reaction between \ce{O2+} (\textit{v} = 0) and the \ce{C3H4} isomers were explored using KinBot to search for barrierless (i.e., submerged relative to the reactants) pathways to products. KinBot has been described in-depth previously,\cite{KinBot2020,KinBot2023} and only a brief description is provided here. KinBot is an open-source chemical kinetics workflow code that can explore and characterize reactive potential energy surfaces typically with the aim to predict reaction rate constants. KinBot calculations begin with a geometry optimization of a specified `reactant' molecule, which in this study is the \ce{[O2-C3H4]+} reaction complex. After this optimization, KinBot generates saddle point guess structures by systematically modifying the initial geometry via a series of constrained optimizations as dictated by reaction templates. These guess structures are then optimized to true saddles points, and, if the optimization is successful and the structure’s energy is below a user-defined threshold, intrinsic reaction coordinate (IRC) calculations are used to ensure the connectivity of the saddle to the reactant and to identify the product(s). If the reaction product is an isomer of the original well, a new exploration is started by the code until all connected and valid wells are found.

KinBot uses various levels of theory in its workflow. The exploration phase is typically done at a low level of theory, which was L1 = UB3LYP/6-31+G(d) in this work. Then, the structures are refined at a higher level of theory to obtain better geometries and rovibrational properties, L2. Finally, if desired, the energies can be further refined using single-point energy corrections, L3. For L2 and L3, we used various methods to better understand the underlying uncertainties for the reaction paths of interest. The reported energies (shown in Supplementary Information Tables 2 and 3) are at UMP2/aug-cc-pVTZ, $\omega$B97X-D/6-311++G(d,p), and CCSD(T)-F12a/cc-pVDZ-F12//UMP2/aug-cc-pVTZ levels of theory. All intermediates and transition states were assumed to be doublets on the \ce{C3H4O2+} PES. \ce{O2+} is assigned D$_{2h}$ symmetry in Molpro 2023,\cite{Molpro2023} which can use only Abelian point groups. The ground state of \ce{O2+} is $^{2}\Pi_{g}$, which translates into B$_{2g}$ or B$_{3g}$ in Molpro. The calculated energies for these two states are degenerate, in line with the underlying D$_{2h}$ symmetry. \ce{C2O+} is also linear but not centrosymmetric (its connectivity is C-C-O), therefore, its Abelian group is C$_{2v}$. The ground state of \ce{C2O+} is the $^{2}\Pi$ state,\cite{Jutier2007} correlating with the degenerate (for this system) B$_{1}$ and B$_{2}$ symmetries. However, unlike for \ce{O2+}, the CCSD(T)-F12 method yields two non-degenerate states, of which we report the lower one. A more accurate calculation would involve a multi-reference method instead of the single-reference CCSD(T)-F12 method, which further requires careful choices of other species for proper referencing of relative energies. To reduce some of the uncertainties introduced by the single-reference method for \ce{C2O+} we report both a CCSD(T)-F12 energy and a value determined using the ATcT v.1.220 database for comparison.\cite{Ruscic2004,Ruscic2005}
For the DFT and MP2 calculations we used Gaussian 16 \cite{Gaussian16}, while for the CCSD(T)-F12 energies we used Molpro.

Using the above workflow, we were able to determine an expansive potential energy surface for this reaction. We found a very large number of pathways, with $>$150 wells and $>$200 bimolecular product channels at L1, all satisfying the energetic criterion, which was that no barrier shall be above the energy of the starting species. The reason for the vastness of the PES is that the reactants initially form very strong bonds at the addition step. After this attachment, the PES descends rapidly by breaking the O–O bond, resulting in several wells on the PES that are 6 eV or deeper relative to the reactants. We analyzed the large reaction network using our postprocessing tool, PESViewer,\cite{PESViewer} which allowed us to find feasible pathways between any two species on the surface and explore some pathways manually.

\begin{acknowledgement}
The authors thank Alicja Karpinska for her contributions to data acquisition. This work was supported by the National Science Foundation (PHY-2317149) and the Air Force Office of Scientific Research (FA9550-16-1-0117). J.Z. is supported by the U.S. Department of Energy (DOE), Office of Science, Office of Basic Energy Sciences, Division of Chemical Sciences, Geosciences and Biosciences via the Gas Phase Chemical Physics program. This article has been authored by employees of National Technology and Engineering Solutions of Sandia, LLC, under Contract No. DE-NA0003525 with the U.S. DOE. The employees co-own right, title, and interest in and to the article and are responsible for its contents. The United States Government retains and the publisher, by accepting the article for publication, acknowledges that the United States Government retains a nonexclusive, paid-up, irrevocable, worldwide license to publish or reproduce the published form of this article or allow others to do so, for United States Government purposes. The DOE will provide public access to these results of federally sponsored research in accordance with the DOE Public Access Plan https://www.osti.gov/public-access.

\end{acknowledgement}

\begin{suppinfo}
Includes additional potential energy surfaces, a table of fitted rate constants, and tables of additional thermochemical calculations (PDF)
\end{suppinfo}

\section{References}

\providecommand{\latin}[1]{#1}
\makeatletter
\providecommand{\doi}
  {\begingroup\let\do\@makeother\dospecials
  \catcode`\{=1 \catcode`\}=2 \doi@aux}
\providecommand{\doi@aux}[1]{\endgroup\texttt{#1}}
\makeatother
\providecommand*\mcitethebibliography{\thebibliography}
\csname @ifundefined\endcsname{endmcitethebibliography}  {\let\endmcitethebibliography\endthebibliography}{}

\end{document}